%% file: sn-article.tex
\RequirePackage{amsthm}
\documentclass[referee,sn-mathphys,Numbered]{sn-jnl}

\usepackage{float}
\usepackage{lscape}
\usepackage{caption}
\usepackage{subcaption}

\usepackage{graphicx}%
\usepackage{multirow}%
\usepackage{amsmath,amssymb,amsfonts}%
\usepackage{amsthm}%
\usepackage{mathrsfs}%
\usepackage[title]{appendix}%
\usepackage{xcolor}%
\usepackage{textcomp}%
\usepackage{manyfoot}%
\usepackage{booktabs}%
\usepackage{algorithm}%
\usepackage{algorithmicx}%
\usepackage{algpseudocode}%
\usepackage{listings}%

\theoremstyle{thmstyleone}%
\theoremstyle{thmstyletwo}%

\theoremstyle{thmstylethree}%

\begin{document}

\title[Magnetic fields in wormholes]{Dynamical black holes in the inflationary epoch}


\author*[1]{\fnm{Milos} \sur{Ertola Urtubey}}\email{meusay@fcaglp.unlp.edu.ar}

\author[1]{\fnm{Daniela} \sur{P\'erez}}



\affil[1]{\orgname{Instituto Argentino de Radioastronom\'ia (IAR, CONICET/CIC/UNLP)},  \city{Villa Elisa}, \postcode{C.C.5, (1894)}, \state{Buenos Aires}, \country{Argentina}}




\abstract{We investigate the evolution of black holes present during the inflationary epoch, assuming they are dynamically coupled to the cosmological background through a generalized McVittie geometry, such that their gravitational mass scales with the cosmic scale factor. Adopting Starobinsky's $\mathcal{R}^2$ inflation model, we analyse the combined effects of cosmological coupling, Hawking evaporation and radiation accretion during the subsequent cosmic eras: inflation, radiation, matter, and dark energy. Requiring the black hole event horizon to remain smaller than the particle horizon at all times yields an upper bound on the mass parameter. Radiation accretion during the radiation era further constrains the parameter space to prevent runaway growth. Hawking evaporation sets a lower bound on the initial mass to ensure survival through inflation. We find that only black holes formed within a narrow initial mass range during inflation can persist to the present day, reaching a maximum mass of $M(t_0) \simeq 1.043\times10^{-3} M_\odot$.}




\maketitle

\section{Introduction}

Black holes are unlike any other objects in the universe; strictly speaking, they are regions of spacetime bounded by an event horizon. This one-way membrane constitutes their defining feature. Within the framework of General Relativity (GR), only three parameters—mass $M$, angular momentum $J$, and charge $Q$—are sufficient to describe them completely. Since black holes, however, are essentially spacetime regions, any modification in the dynamics of the background spacetime may affect their properties.

Since the seminal work of McVittie in 1933 \cite{mcv33}, there has been intense research on the coupling between cosmological background dynamics and local objects, in particular black holes (see, for instance, Refs.~\cite{car+10a,car+10b,nan+12a,nan12b,far15,com+19,com+20,spe+22}). It is worth noting that the classical black hole solutions in GR are asymptotically flat; therefore, these metrics are not adequate to describe black holes embedded in a dynamical cosmological background. In fact, it was recently shown \cite{far+24} that embedding an exactly static black hole horizon in a time-dependent geometry gives rise to a naked null singularity. From a theoretical standpoint, the latter implies that black hole event horizons embedded in an FLRW geometry must evolve over time. There exist exact solutions to the Einstein field equations that represent cosmological black holes, namely black holes coupled to the background cosmological dynamics. Some well-known spacetime metrics describing such objects include the McVittie spacetime \cite{mcv33,kal+10,lak+11,nol98,nol99a,nol99b,nol17} and its generalized version \cite{Faraoni-2007,far15}, the Sultana--Dyer solution \cite{sul+05}, the Culetu metric \cite{cul12,sat+22}, and cosmological coupled black holes with regular horizons \cite{cad+26}, and regular coupled cosmological black holes \cite{cad+23,cad+25}.

It has been shown that, in the case of the generalized McVittie spacetime \cite{car+10b,per+22}, the gravitational energy of the black hole is proportional to the background scale factor. As a consequence, the black hole event horizon may either shrink or grow in response to the cosmic evolution of the universe. This property of cosmological black holes could be relevant in certain astrophysical contexts. On the one hand, the effects of the global dynamics may become significant when the cosmological background evolves rapidly, as during the inflationary epoch. On the other hand, such effects could also be important for black holes that have existed over timescales comparable to the age of the universe.

The goal of this work is to investigate the evolution of black holes that were present during the inflationary period and have survived to the present epoch. We assume that these black holes are coupled to the background cosmological dynamics and that, as a result, their properties are modified throughout the different cosmic eras. We derive mass constraints on these cosmological black holes by taking into account both accretion from the surrounding medium and mass loss due to Hawking radiation.

Black holes may be present during the inflationary epoch within different, physically well-motivated theoretical frameworks. Since inflation is well approximated by a (quasi) de Sitter phase, semiclassical gravity allows for the nucleation of black holes through nonperturbative processes, such as black hole pair creation, leading to the existence of already formed black holes during inflation \cite{gin+83,bou+96}. In addition, in cosmological scenarios extending beyond the standard Big Bang, such as bouncing models, black holes may originate in a pre-inflationary phase and subsequently persist through both the bounce and the inflationary expansion \cite{per+22}. In this work, we therefore consider the evolution of black holes present during inflation, either as objects formed during this epoch or as relics from an earlier cosmological phase, and investigate their coupling to the background cosmological dynamics.

In the next section, we present the main properties of the cosmological black hole solution we employ in this work.


\section{Dynamical black holes and cosmological coupling}

Faraoni and Jacques originally proposed the generalized McVittie spacetime \cite{Faraoni-2007,far15}; the line element in isotropic coordinates $(t,r,\theta,\phi)$ takes the form:

\begin{equation}\label{dsGMV}
ds^{2}  =  -\frac{\left[1 - \frac{G\;  m(t)}{2 c^2 r}\right]^2}{\left[1 + \frac{G \; m(t)}{2 c^2 r}\right]^2} c^2 dt^2 + a^2(t) \left[1 + \frac{G \; m(t)}{2 c^2 r}\right]^4 \left[dr^2 + r^{2} \ \sin^{2}{\theta}\right].
\end{equation}

In the expression above, $a(t)$ is the scale factor of the cosmological background and $m(t)$ denotes a function that depends on the cosmic time $t$.

The source of this geometry is an imperfect fluid with energy-momentum tensor described by
\begin{equation}\label{em}
T_{ab} = \left(\frac{p}{c^2} + \rho\right) u_a u_b+ p \; g_{ab} + q_a \;u_b + q_b \;u_a.
\end{equation}
As usual, $\rho$ and $p$ stand for the density and pressure of the fluid, respectively; we assume that the four-velocity $u^{a}$ has components
\begin{equation}
u^{\mu} = \left(\frac{\left[1 + \frac{G \; m(t)}{2 c^2 r}\right]}{\left[1 - \frac{G\;  m(t)}{2 c^2 r}\right]},0,0,0\right),
\end{equation}
while the purely spatial vector field $q^{a}$ represents the current density of heat
\begin{equation}
q^{\alpha} = (0,q,0,0), \;\;\; q^{b} u_{b} = 0.
\end{equation}

If we substitute the metric coefficients given by \eqref{dsGMV} and the energy-momentum tensor \eqref{em} into Einstein field equations, it is possible to derive expressions for $\rho(t,r)$, $p(t,r)$ and $q(t,r)$ in terms of $a(t)$ and $m(t)$. In particular, the $(0,1)$ component of Einstein field equations is

\begin{equation}
\begin{split}
\frac{\dot{a}(t)}{a(t)}+\frac{\dot{m}(t)}{m(t)} =- 4\pi r^2 \frac{a^2(t)}{m(t)}\left[1 + \frac{G \; m(t)}{2 c^2 r}\right]^2 \left[1 - \frac{G\;  m(t)}{2 c^2 r}\right]^4 q.
\end{split}
\end{equation}

We see that if $q = 0$, the energy-momentum tensor corresponds to a perfect fluid. The solution of the later equation is
\begin{equation}\label{sol_mcvittie}
m(t) = \frac{m_0}{a(t)},
\end{equation}
being $m_0$ a non-negative constant. If we replace \eqref{sol_mcvittie} into \eqref{dsGMV}, we get the McVittie metric. As discussed by Carrera and Giulini \cite{car+10a}, in the newtonian limit the constant $m_0$ represents the mass of the central object.

In this work, we focus on a particular class of generalized McVittie solutions that correspond to the choice
\begin{equation}\label{comoving}
m(t) = m_0,
\end{equation}
where $m_0$ is a constant.



In GR, for spherically symmetric spacetimes, the Misner--Sharp--Hernandez (MSH) mass provides a quasi-local measure of the mass (or energy) contained within a spherical region. The MSH mass can be defined geometrically in terms of the Riemann tensor and admits a natural decomposition into two contributions: one associated with the Ricci curvature and another associated with the Weyl curvature. The Ricci part of the MSH mass can be locally related to the energy--momentum tensor of the matter fields, while the gravitational mass of the central object is encoded in the Weyl part \cite{car+10b}. Owing to its quasi-local and covariant definition, the MSH mass provides a well-defined notion of black hole mass in expanding universes, where global definitions of mass based on asymptotic flatness are no longer applicable. In particular, it represents the relativistically correct measure of the total gravitating energy enclosed within a spherical surface of areal radius $R$, remaining valid even in fully dynamical and cosmological spacetimes \footnote{In Ref. \cite{cad+24}, it is argued that the MSH mass offers a direct measure of the possible cosmological coupling of local objects embedded a in cosmological background.}.

The Weyl part of the MSH mass for the generalized McVittie spacetime under the choice \eqref{comoving} is \cite{car+10b}
\begin{equation}
E_{\mathrm{W}} = m_0 \; a(t).
\end{equation}
We interpret this quantity as the effective gravitational mass of the central object. Its time dependence does not correspond to conventional matter accretion but rather to the dynamical embedding of the black hole within an expanding background. Thus, in what follows, we identify the gravitational mass of the black hole with the Weyl part of the MSH mass, that is,
\begin{equation}
M(t) = m_0 \; a(t).
\end{equation}

Furthermore, as shown in Ref. \cite{per+22}, in terms of the areal radius coordinate $R$, the radius of the surface that encloses the black hole as described by this class of generalized McVittie solutions is
\begin{equation}
R_\mathrm{BH}(t) = 2 G \frac{m_0 \; a(t)}{c^2}.
\end{equation}

The evolution $M(t) \propto a(t)$ therefore reflects the dynamical response of the black hole horizon to the background expansion, rather than a coordinate artifact. The black hole is coupled to the background dynamics; we see that its size changes according to the cosmic evolution of the universe.


If black holes were present during the inflationary period, the size of these objects could, in principle, grow exponentially and eventually surpass the particle horizon. However, this cannot be the case, since we do not appear to be living inside a black hole \cite{1972Natur.240..298P,2023mgm..conf.1327P}. It is therefore expected that certain conditions must be satisfied—such as constraints on the black hole mass—to ensure that the event horizon remains at all times within the particle horizon. As we will demonstrate in the following sections, these requirements place nontrivial bounds on the properties of black holes embedded in an inflationary background.

In order to analyze the evolution of both horizons, the model of cosmological inflation to be considered must first be specified.

\section{Starobinsky $\mathcal{R}^2$ inflation}

Inflation is an early period of accelerated expansion in the universe. It was introduced to solve many problems presented by the standard cosmological model, such as the flatness, horizon, and superhorizon correlation problems.

Many of the models considered for this epoch are based on slow-roll inflation, in which inflation is driven by a scalar field that initially takes large values and gradually rolls down a nearly flat potential, marking the end of the inflationary epoch.

One model that best adjusts to Planck measurements \cite{2020Planck} is Starobinsky $\mathcal{R}^2$ inflation \cite{sta80}.

\begin{equation}
    S=\frac{1}{2\kappa^2}\int d^4x\sqrt{-g}f(R)+\int d^4x\mathcal{L}_{M}(g_{\mu\nu},\Psi_{M}),
\end{equation}
where $\kappa^2=8\pi G$, $g$ is the determinant of the metric $g_{\mu\nu}$, and $\mathcal{L}_{M}$ is a matter Lagrangian that depends on $g_{\mu\nu}$ and matter fields $\Psi_{M}$, with $f(R)=R+R^2/(6M)$; the constant $M$ has a dimension of mass, and $R$ denotes the Ricci scalar.

It is known that for an action with a non-linear function $f(R)$, an action in the Einstein frame can be derived under a conformal transformation $\tilde{g}_{\mu\nu}=\Omega^2 g_{\mu\nu}$, where $\Omega^2$ is the conformal factor and a tilde represents quantities in the Einstein frame. Thus, the action in the Einstein frame can be expressed as \cite{def+10}

\begin{equation}
    S_{E}=\int d^4x\sqrt{-\tilde{g}}\left(\frac{1}{2\kappa^2}\tilde{R}-\frac{1}{2}\tilde{g}^{\mu\nu}\partial_{\mu}\phi\partial_\nu\phi-V(\phi)\right)+\int d^4x\mathcal{L}_{M}(g_{\mu\nu},\Psi_{M}),
\end{equation}

where $V(\phi)=\left(F(R)R-f(R)\right)/\left(2\kappa^2F(R)^2\right)$, with $F(R)=f'(R)$.

An approximate solution to the cosmological equations in $f(R)$-theory for the Starobinsky's model takes the form (see \cite{def+10} for the detailed calculations)
\begin{equation}
    \begin{split}
        H&\simeq H_{\mathrm{i}}-\frac{M^2}{6}(t-t_{\mathrm{i}}),\\
        a&\simeq a_{\mathrm{i}}\exp{[H_{\mathrm{i}}(t-t_{\mathrm{i}})-\left(M^2/12\right)(t-t_{\mathrm{i}})^2]},\\
        R&\simeq 12H^2-M^2,
    \end{split}
\end{equation}
where $H_{\mathrm{i}}$ and $a_{\mathrm{i}}$ denote the Hubble parameter and the scale factor evaluated at $t = t_{\mathrm{i}}$, corresponding to the onset of inflation.

The mathematical condition that characterizes the end of the inflation period is
\begin{equation}
\epsilon_{\mathrm{f}} = - \frac{\dot{H_{\mathrm{f}}}}{H^2_{\mathrm{f}}} \simeq \frac{M^2}{6 H^2_{\mathrm{f}}} \simeq 1,\\\ \Rightarrow \\\     H_{\mathrm{f}} \simeq \frac{M}{\sqrt{6}}.
\end{equation}
The quantity $\epsilon_{\mathrm{f}}$ is the slow-roll parameter at the end of inflation 
($t_{\mathrm{f}}\simeq t_{\mathrm{i}}+6H_{\mathrm{i}}/M^2$), and $H_{\mathrm{f}} \equiv H(t_{\mathrm{f}})$. The number of e-foldings is \cite{def+10}
\begin{equation}
    N\simeq \frac{H_{\mathrm{i}}^2}{M^2}
\end{equation}

In the Einstein frame, these quantities correspond to

\begin{equation}
    \begin{split}
        \tilde{a}(\tilde{t})&\simeq\left(1-\frac{M^2}{12H_{\mathrm{i}}^2}M\tilde{t}\right)\tilde{a}_{\mathrm{i}}e^{M\tilde{t}/2},\\
        \tilde{H}(\tilde{t})&\simeq \frac{M}{2}\left[1-\frac{M^2}{6H_{\mathrm{i}}^2}\left(1-\frac{M^2}{12H_{\mathrm{i}}^2}M\tilde{t}\right)^{-2}\right],\\
        V\left(\phi\right)&\simeq\frac{3M^2}{4\kappa^2}\left(1-2e^{-\sqrt{2/3}\kappa\phi}\right)^2
    \end{split}
\end{equation}

Having obtained a scale factor that describes an inflationary epoch within General Relativity, we can now model the evolution of the scale factor from the beginning of inflation to the present day, incorporating the successive cosmological epochs of radiation, matter, and dark energy field:
\begin{equation}
    a(t)= \begin{cases} 
      \left(1-\frac{M^2}{12H_{\mathrm{i}}^2}Mt\right)a_{\mathrm{inf}}\,e^{Mt/2}, & t_{\mathrm{i}}\leq t \leq t_{\mathrm{f}} \\
      a_{\mathrm{rad}}\,t^{1/2}, & t_{\mathrm{f}}\leq t\leq t_{\mathrm{rm}} \\
      a_{\mathrm{mat}}\,t^{2/3}, & t_{\mathrm{rm}}\leq t\leq t_{\mathrm{mde}}\\
      e^{H_{0}(t-t_{0})}, & t_{\mathrm{mde}}\leq t
   \end{cases}
\label{scalefactor}
\end{equation}
where $t_{\mathrm{rm}}$ denotes the time of the radiation-matter transition, $t_{\mathrm{mde}}$ the matter-dark energy transition, $t_{0}$ as today, with $a(t_{0})=1$; the terms $a_{\mathrm{inf}},\,a_{\mathrm{rad}},\,a_{\mathrm{mat}}$ are such that the scale factor is continuous. We show in Table \ref{table:Table1} the numerical values of these quantities used in this work.

It is important to note that the start of inflation $t=t_\mathrm i=0$ does not coincide with the essential singularity; in the Einstein frame the singularity is reached at some time $t<0$.
\begin{table}
\centering
\begin{tabular}{ccc}
\hline\hline\noalign{\smallskip}
$t_{\mathrm{i}}$ [$s$] & Onset of Inflation & 0 \\ 
$t_{\mathrm{f}}$ [$s$] & End of Inflation & $9.21497\times10^{-36}$ \\
$t_{\mathrm{rm}}$ [$s$] & Radiation-Matter Transition & $1.47321\times10^{12}$ \\ 
$t_{\mathrm{mde}}$ [$s$] & Matter-Dark Energy Transition & $3.09264\times10^{17}$ \\ 
$t_{0}$ [$s$] & Today Time & $4.35197\times10^{17}$ \\ 
$M$ [$s^{-1}$] & Parameter Starobinsky $\mathcal{R}^2$ inflation & $1.51927\times10^{-37}$ \\
$H_{\mathrm{i}}$ [$s^{-1}$] & Hubble Parameter at $t=t_\mathrm i$ & $7.33876\times10^{-37}$ \\
$H_{0}$ [$s^{-1}$] & Hubble Parameter at $t=t_0$ &  $2.58199\times10^{-18}$\\
$a_{\mathrm{inf}}$ & & $4.07365\times10^{-58}$ \\
$a_{\mathrm{rad}}$ & & $1.6878\times10^{-10}$ \\
$a_{\mathrm{mat}}$ & & $1.58047\times10^{-12}$ \\
\hline
\end{tabular}
\caption{Numerical values of different cosmological parameters adopted in this work.}
\label{table:Table1}
\end{table}


\subsection{Black hole horizon evolution: Comparison with the particle horizon}

Let us consider a cosmological black hole present in the inflationary period at time $t_{\mathrm{i}}$. As discussed above, the gravitational mass of the black hole changes over time due to the coupling with the background dynamics as
\begin{equation}
    M(t)= m_{0} a(t),
\end{equation}
and the black hole horizon is defined as
\begin{equation}
    R_{\mathrm{BH}}(t)=2 \frac{G M(t)}{c^2}=2 \frac{G m_0}{c^2}a(t).
\end{equation}

For clarity of notation, we write $m_{0} = \gamma\, w$, where $w = 1\,\mathrm{g}$ and $\gamma$ is a dimensionless quantity that we refer to as the mass parameter. In the following sections, we determine the allowed values of this parameter.

An additional relevant quantity is the particle horizon, denoted by $d_{\mathrm{h}}$, defined as the physical distance traveled by a light signal since the beginning of the cosmological evolution\cite{2022Baumann}
\begin{equation}
d_{\mathrm{h}}(t)=a(t)\int_{t_{\mathrm{i}}}^{t}\frac{c\,dt'}{a(t')}.
\label{2.8}
\end{equation}
This quantity measures the size of the region that is in causal contact with a comoving observer at time $t$.

Since the Universe as a whole is not a black hole \cite{1972Natur.240..298P,2023mgm..conf.1327P}, we require that the growth  of the black hole event horizon remain smaller than that of the particle horizon. This condition enables us to estimate an upper bound on the black hole mass during the inflationary epoch. More precisely, we impose that the increase in the event horizon radius from the onset to the end of inflation be smaller than the corresponding increase in the particle horizon over the same period. This yields:

\begin{eqnarray}
\Delta R_{\mathrm{BH}}=R_{\mathrm{BH}}(t_{\mathrm f})-R_{\mathrm {BH}}(t_{\mathrm i})&\leq &d_{\mathrm h}(t_\mathrm f),\\
    \frac{2G \gamma w}{c^2}(a(t_{\mathrm{f}})-a(t_{\mathrm{i}}))&\leq &d_{\mathrm{h}}(t_{\mathrm{f}}),\\
    \gamma&\leq &\gamma_1=\frac{c^2}{2G w \left(a(t_\mathrm f)-a(t_\mathrm i)\right)}d_\mathrm h(t_\mathrm f).
\end{eqnarray}

We obtain $\gamma_{1}=6.34126\times10^{58}$. Therefore, if a black hole has a mass greater than $M_{\mathrm{i}}=M(t_\mathrm i)=\gamma_{1}wa(t_{\mathrm{i}})=26.7644 \,\text{g}$ at the start of inflation, the black hole's horizon grows faster than the particle horizon, and it will not be considered in our analysis.

Note that since $R_{\mathrm{BH}}$ and $d_\mathrm h(t)$ are monotonic during inflation, ensuring that the increase of the black hole radius remains smaller than the particle horizon at the end of inflation guarantees that $R_{\mathrm{BH}}(t) < d_\mathrm h(t)$  at all intermediate times.

In the following cosmological epoch, radiation, the event horizon must remain smaller than the particle horizon, namely, $R_\mathrm {BH}(t)\leq d_\mathrm h(t_\mathrm f)$. Assuming that $\gamma\leq\gamma_1$, we derive a new upper limit for $\gamma$ 
\begin{equation}
    R_{\mathrm{BH}}(t)\leq d_{\mathrm{h}}(t), \forall t,\,t_{\mathrm{f}}\leq t\leq t_{\mathrm{rm}}.
\end{equation}
The latter inequality is satisfied if the mass parameter $\gamma$, $\gamma\leq\gamma_{2}=1.3324\times 10^{45}$.

Following the discussion in \cite{car+21}, a black hole is typically expected to have a mass no smaller than the Planck mass, $M_{\mathrm{BH}} \sim m_{\mathrm{Planck}} = 2.17645 \times 10^{-5}\,\mathrm{g}$, below which the classical description is no longer reliable. Therefore, for the mass parameter $\gamma_2$, the black hole cannot form earlier than $t_{\mathrm{formation}} = 2.31758 \times 10^{-36}\,\mathrm{s}$, since at earlier times its mass would lie below this threshold. By the end of inflation, the black hole mass has grown to $M(t_{\mathrm{rm}}) = 7.07744 \times 10^{17}\,\mathrm{g} = 3.55829 \times 10^{-16}\; M_{\odot}$.

\begin{figure}[t]
    \centering
    \includegraphics[width=0.7\textwidth] {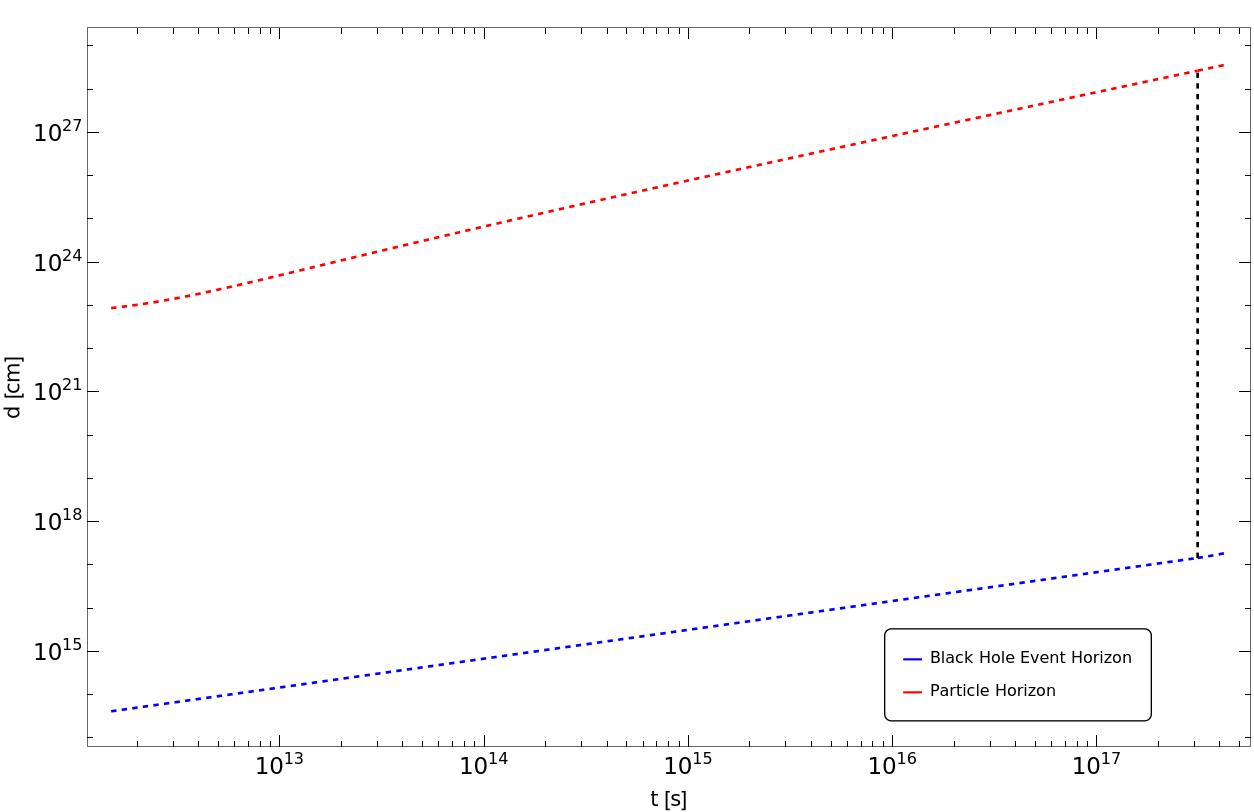}
    \caption{Time evolution of the black hole event horizon and the particle horizon on a logarithmic scale during the matter- and dark energy–dominated epochs, for $\gamma=\gamma_2$. The vertical dashed line marks the transition between the two cosmological regimes.}
  \label{fig:Horizons}
\end{figure}

In the subsequent cosmological epochs, no additional constraints need to be imposed on the parameter $\gamma$, since the particle horizon always remains larger than the event horizon, as illustrated in Fig.~\ref{fig:Horizons}.

\section{Evolution of the gravitational mass} 

In our previous analysis, we found that if a black hole is present at the inflationary epoch at time $t_{\mathrm{i}}$, there is an upper limit of its initial gravitational mass such that the particle horizon always remains larger than the event horizon at all cosmic epochs. 
This upper limit, however, does not take into account that after the inflationary epoch, the mass of the black hole, and hence its event horizon, could change because of other processes. The mass of the black hole also grows because of accretion of the cosmological fluid and decreases due to Hawking radiation. This non-linear evolution of the mass could place further constraints on the initial mass of the black hole at the inflationary epoch.



We begin the analysis in the radiation epoch. The equation that governs mass evolution is:
\begin{equation}\label{eq-tot}
\left.\frac{dM}{dt} \right|_{\mathrm{tot}}
= \left.\frac{dM}{dt} \right|_{\mathrm{cos}} + \left.\frac{dM}{dt} \right|_{\mathrm{rad}} + \left.\frac{dM}{dt} \right|_{\mathrm{accr}},
\end{equation}
where each term on the right-hand side corresponds to:
\begin{itemize}
\item Cosmological coupling.
\begin{equation}
\left.\frac{dM}{dt} \right|_{\mathrm{cos}} = \gamma w\dot{a}(t).
\end{equation}
\item Hawking radiation \cite{haw75}.
\begin{equation}
\left.\frac{dM}{dt} \right|_{\mathrm{rad}} = -\frac{A(M)}{M(t)^2},
\end{equation}
where

\begin{equation}
    A(M)= \begin{cases} 
      7.8\,\times\,10^{27}\, \text{g}^3\,\text{s}^{-1}, & M\leq 10^{15}\, \text{g} \\
      7.87825\times10^{27}\,\text{g}^3\,\text{s}^{-1} - 7.82525\times10^{10}\,\text{g}^2\,\text{s}^{-1} M, & 10^{15}\,\text{g}<M< 10^{17}\, \text{g} \\
      5.3\,\times\,10^{25}\, \text{g}^3\,\text{s}^{-1}, & 10^{17}\, \text{g}\leq M
   \end{cases}
\end{equation}

\item Accretion of radiation fluid \cite{zel+67}.
\begin{equation}\label{acr}
\left.\frac{dM}{dt} \right|_{\mathrm{accr}} = \frac{27\pi G^2}{c^3}\rho_{\mathrm{rad}}(t)M(t)^2,
\end{equation}
and
\begin{equation}\label{rho-rad}
    \rho_{\mathrm{rad}}(t)=\frac{3}{8\pi G}H(t)^2=\frac{3}{8\pi G}\left(\frac{\dot{a(t)}}{a(t)}\right)^2.
\end{equation}
\end{itemize}

We emphasize that the accretion rate given in Eq.  \eqref{acr} to the effective Zel’dovich–Novikov prescription derived for a Schwarzschild black hole immersed in a thermal bath. In the present context, however, the black hole is described by a dynamical cosmological solution rather than an asymptotically flat Schwarzschild geometry. Therefore, Eq. \eqref{acr} should be understood as a phenomenological estimate of the maximal impact of radiation accretion on the mass evolution. Our goal is not to provide a fully self-consistent relativistic treatment of accretion onto a cosmological black hole, but rather to explore the mass ranges for which such objects can survive the early universe under simplified and controlled assumptions. A complete description would require modeling accretion directly within the generalized McVittie geometry, consistently accounting for the dynamical horizon and background expansion, which lies beyond the scope of this work.












The equation governing the evolution of the gravitational mass during the radiation-dominated era admits only numerical solutions. Before addressing its numerical integration, we first examine the third term, which accounts for the accretion of radiation. Substituting Eq.~\eqref{rho-rad} into Eq.~\eqref{acr}, and using the expression for the scale factor in the radiation era (see Eq.~\eqref{scalefactor}), we obtain
\begin{equation}
    \frac{dM}{dt}=\frac{81 G}{32c^3}\frac{M(t)}{t^2}.
\end{equation}
This equation has an analytical solution given by
\begin{equation}
    \frac{1}{M(t)}=\frac{1}{M(t_{\mathrm{f}})}+\frac{81 G}{32c^3}\left(\frac{1}{t}-\frac{1}{t_{\mathrm{f}}}\right),
\end{equation}
where $t_{\mathrm{f}}$ is the value of the cosmic time at the end of inflation, and $M(t_{\mathrm{f}})$ is the value of the gravitational mass of the black hole at $t_{\mathrm{f}}$.  Depending on the value of $M(t_{\mathrm{f}})$, it may occur that, for some $t_* \in (t_{\mathrm{f}}, t_{\mathrm{rm}}]$, the mass diverges, $M(t_*) \to \infty$, so that the gravitational mass is ill-defined. This motivates the question of determining the upper bound on $M(t_{\mathrm{f}})$ for which the mass diverges by the end of the radiation-dominated era, i.e., such that $M(t_{\mathrm{rm}}) \to \infty$. This critical value is given by
\begin{equation}
    M(t_{\mathrm{f}})\geq \frac{t_{\mathrm{rm}}t_{\mathrm{f}}}{t_{\mathrm{rm}}-t_{\mathrm{f}}}\frac{32c^3}{81G}.
\end{equation}
Therefore, if we only consider that the mass of the black hole changes because of accretion of radiation, there is an upper limit on the value of the mass at the end of inflation that is given by $M(t_{\mathrm{f}})=t_{\mathrm{rm}}t_{\mathrm{f}}/(t_{\mathrm{rm}}-t_{\mathrm{f}})(32c^3)/(81G)\approx1469.65\, \text{g}$. 

When Eq.~\eqref{eq-tot} is solved numerically, the singular behavior induced by the accretion term is inherited by the total mass-evolution equation. To avoid numerical instabilities, we derive an upper bound on the values of the mass parameter $\gamma$: $\gamma_3=2.07348\times10^{30}$, that translates to a mass at the end of inflation of $M(t_{\mathrm{f}}) =1062.35\,\text{g} \simeq 5.34 \times 10^{-31}\, M_{\odot}$.

\begin{figure}[t]
    \centering
    \includegraphics[width=0.7\textwidth] {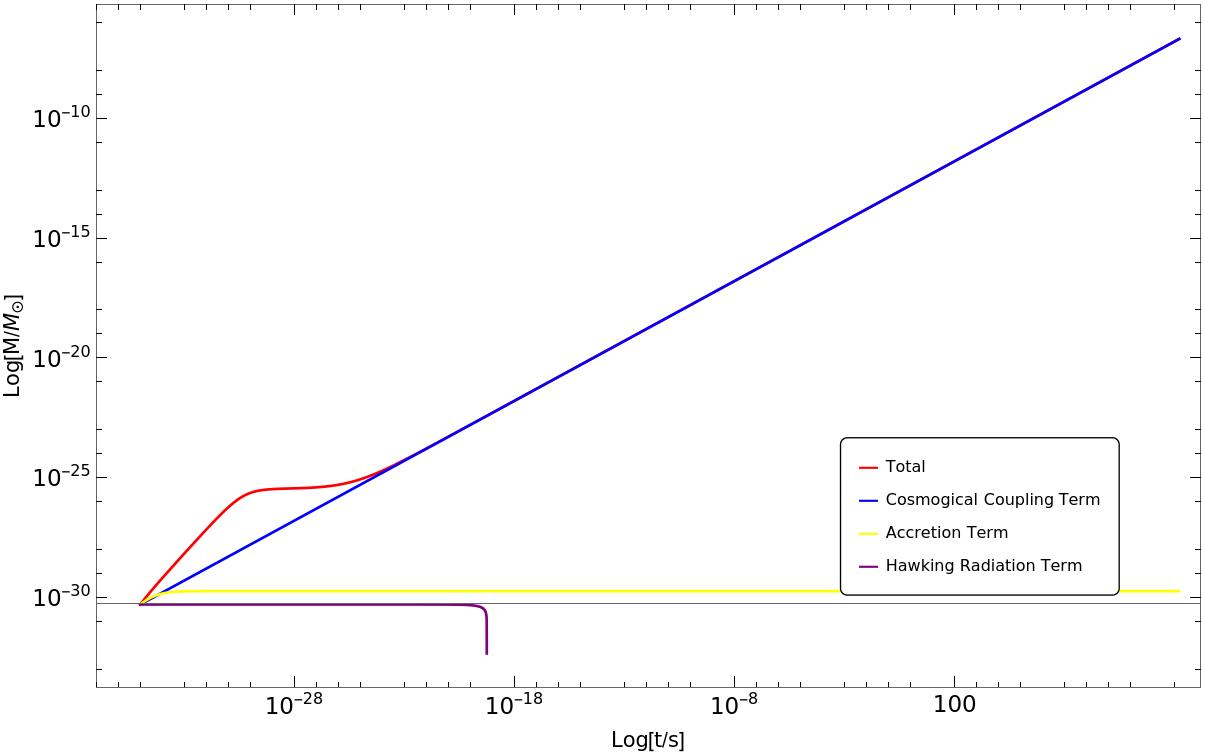}
    \caption{Mass evolution in the radiation era on a logarithmic scale. The black hole has a mass of $M=1062.35\,\text{g}$ at the end of the inflationary period, and a mass of $M(t_{\rm rm})=2.14391\times10^{-7}\,M_{\odot}$ at the end of the radiation era.}
  \label{fig:Radiation}
\end{figure}

\begin{figure}[t]
    \centering
    \includegraphics[width=0.7\textwidth] {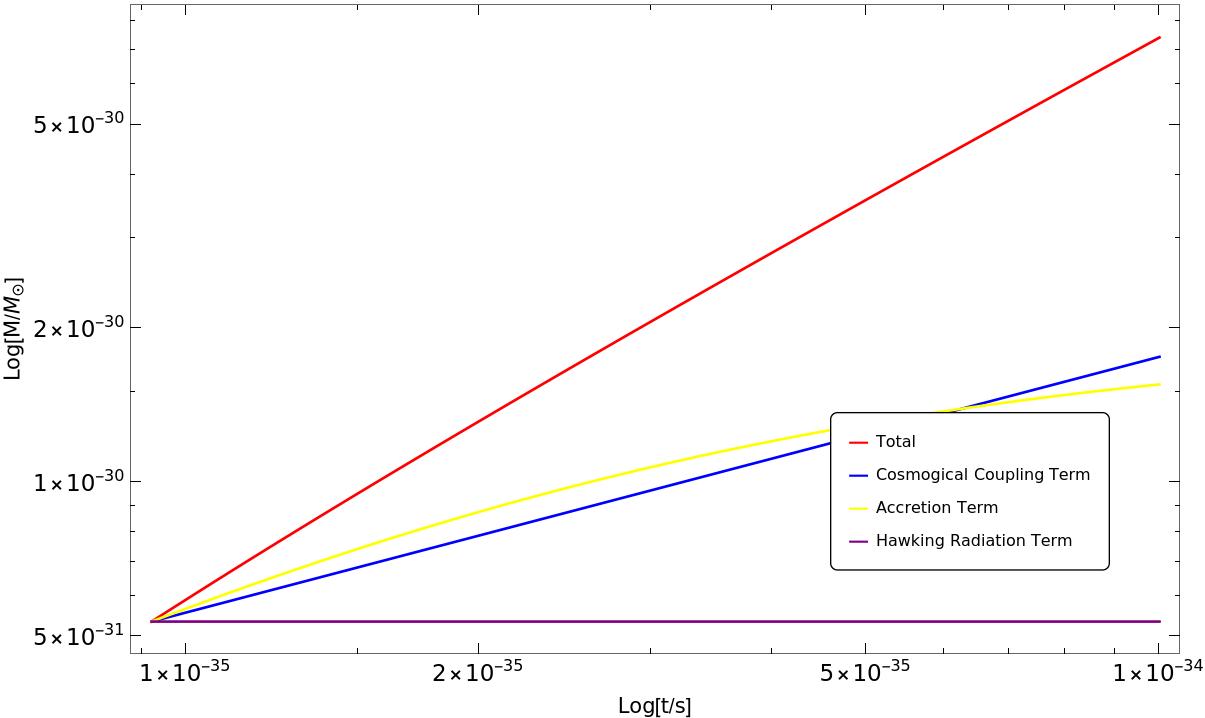}
    \caption{Zoom-in of the early stages of the radiation-dominated epoch shown in Fig.~\ref{fig:Radiation}. At early times, the accretion term clearly dominates the evolution.}
  \label{fig:Zoom}
\end{figure}

We show in Figure \ref{fig:Radiation} the evolution of the gravitational mass of the black hole during the radiation era considering that $M(t_{\mathrm{f}}) = 1062.35\, \text{g}$ and $\gamma=\gamma_3$ at the end of the inflationary period. At the end of the radiation era, the mass is of $M(t_{\mathrm{rm}}) = 2.14391\times10^{-7}\,M_{\odot}$. The plot shown in Figure \ref{fig:Radiation} was obtained by solving numerically Eq. \eqref{eq-tot}. We also show in the plot each of the individual contributions. From this plot is visible that if the Hawking radiation term is the only one considered, then the black hole vanishes in a finite time $t_{\rm dis}\simeq5\times10^{-20}\,\text{s}$. The bump observed in the mass evolution, when all contributions are taken into account, can be explained as follows: At the beginning of the radiation-dominated epoch, the accretion term dominates the cosmological contribution (this can be seen in Fig. \ref{fig:Zoom}), resulting in accelerated mass growth. Over time, however, both the accretion and Hawking radiation terms become negligible and the dynamics are effectively governed by the cosmological coupling alone. Consequently, the solution asymptotically approaches that obtained when only the cosmological coupling is considered.

Note that it is crucial to account for the contributions to mass growth arising both from the coupling to the cosmological dynamics and from radiation accretion. If both effects were neglected, a black hole with an initial mass of $M(t_{\mathrm{f}})=1062.35\,\mathrm{g}$ would not survive the radiation-dominated era, as it would completely evaporate at $t = 5.12 \times 10^{-12}\,\mathrm{s}$.

In summary, we have derived an upper bound on the black hole mass at the end of the inflationary epoch such that (i) the event horizon always remains smaller than the particle horizon, and (ii) the evolution of the black hole mass remains well defined at all times.

Taking this upper bound into account: (i) what is the initial mass of the black hole at the onset of inflation? and (ii) what would be the mass of such black holes at the present epoch?

\begin{figure}[t]
    \centering
    \includegraphics[width=0.7\textwidth] {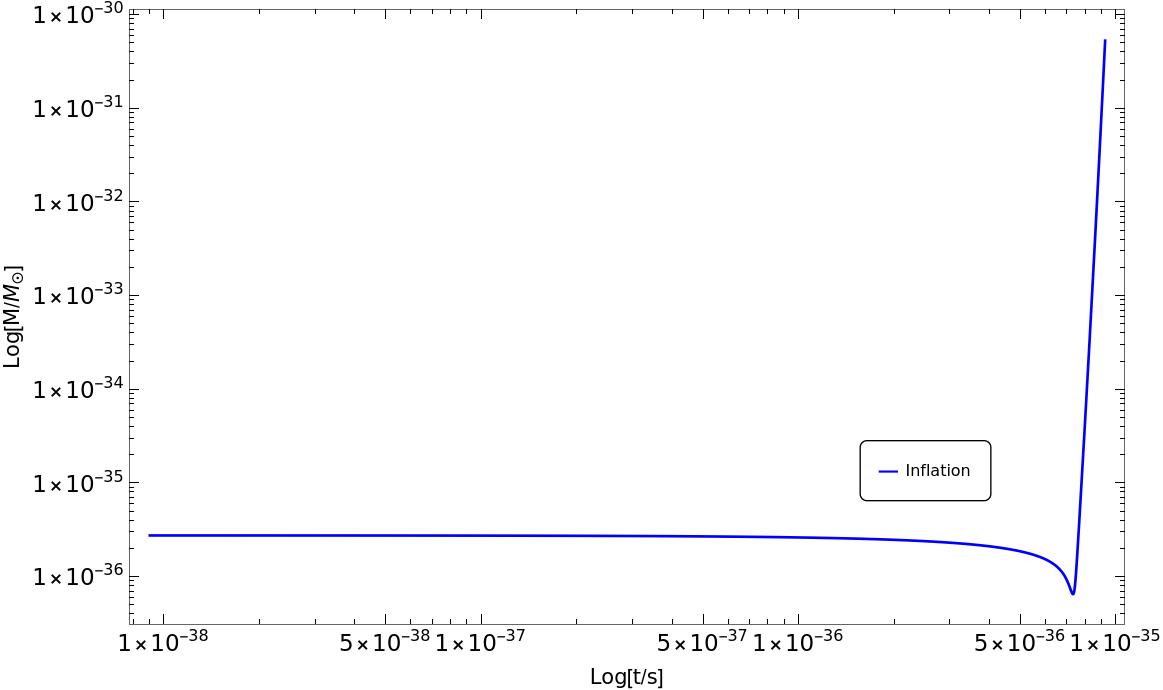}
    \caption{Mass evolution in the inflation era on a logarithmic scale. The black hole has a mass parameter of $\gamma=\gamma_{3}$ at the end of the inflationary period.}
  \label{fig:Infl}
\end{figure}

The answer to the first question is illustrated in Fig.~\ref{fig:Infl}. We solve Eq.~\eqref{eq-tot} while neglecting the accretion term, and integrate the equation backward in time, imposing the final condition $M(t_{\mathrm{f}})=1062.35\,\mathrm{g}$. At the onset of inflation, $t_{\mathrm{i}}=0$, the black hole mass is $M(t_{\mathrm{i}})=5.52687 \times 10^{-3}\,\mathrm{g} = 2.78 \times 10^{-36}\,M_{\odot}$. The coupling between the black hole and the background cosmological dynamics counterbalances the mass loss due to black hole evaporation.

\begin{figure}[t]
    \centering
    \includegraphics[width=0.7\textwidth] {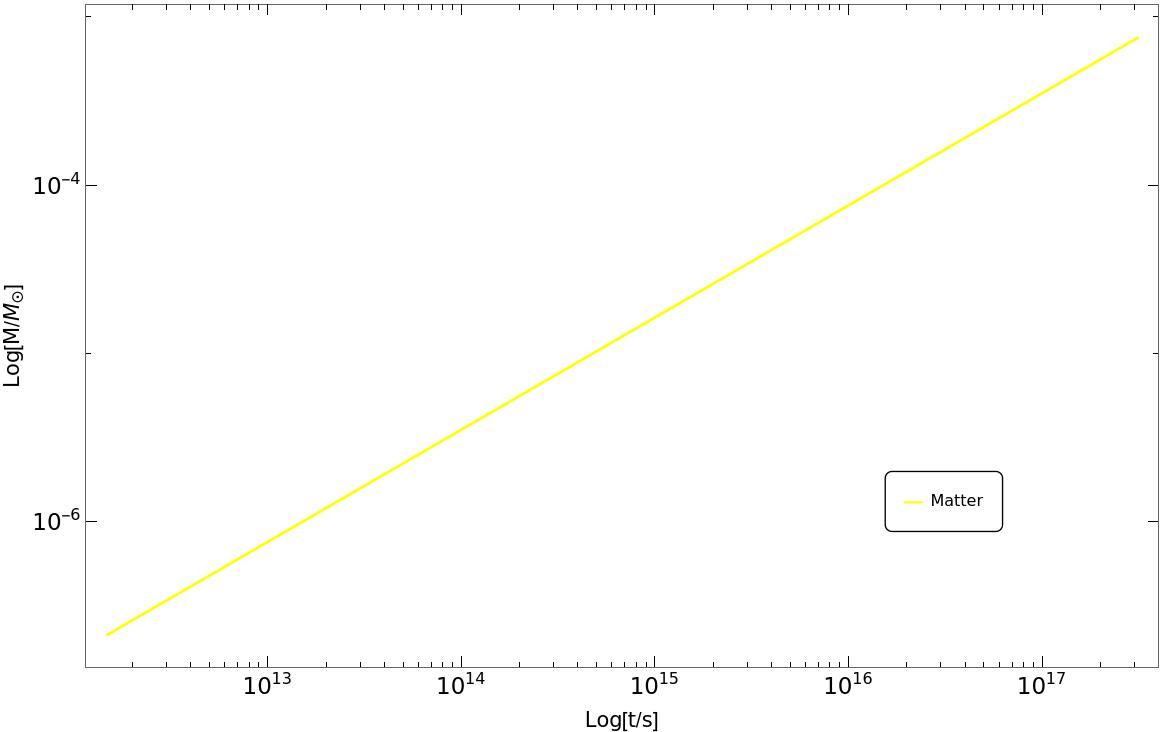}
    \caption{Mass evolution in the matter era on a logarithmic scale. The black hole has a mass of $M(t_{\mathrm{f}})=1062.35\,\text{g}$ at the end of the inflationary period, and ends the matter era with a mass of $M(t_{\rm mdm})=7.53862\times10^{-4}\,M_{\odot}$.}
  \label{fig:Matter}
\end{figure}

\begin{figure}[t]
    \centering
    \includegraphics[width=0.7\textwidth] {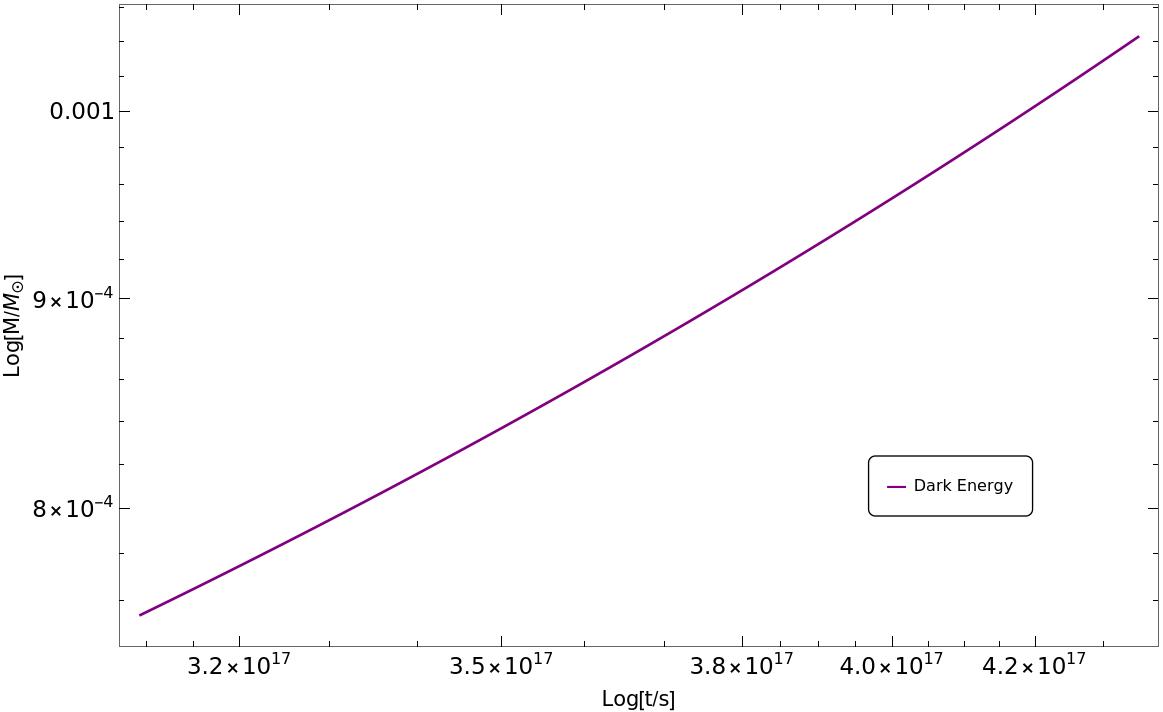}
    \caption{Mass evolution in the dark energy era on a logarithmic scale. The black hole has a mass of $M=1062.35\,\text{g}$ at the end of the inflationary period, and ends the dark energy era with a mass of $M(t_{0})=1.043\times10^{-3}\,M_{\odot}$.}
  \label{fig:DarkEnergy}
\end{figure}

\begin{figure}[t]
    \centering
    \includegraphics[width=0.7\textwidth] {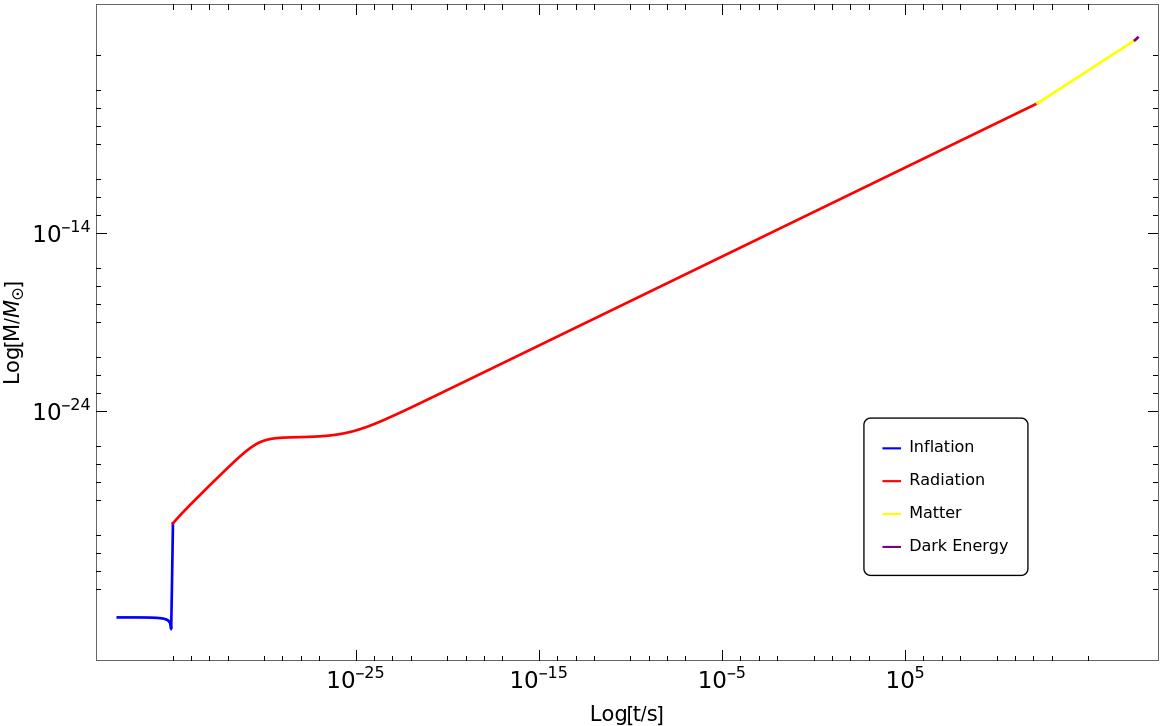}
    \caption{Mass evolution in all the eras on a logarithmic scale. The black hole has a mass of $M=1062.35\,\text{g}$ at the end of the inflationary period, and a mass of $M(t_{0})=1.043\times10^{-3}\,M_{\odot}$ today.}
  \label{fig:History}
\end{figure}

To estimate the black hole mass at the present epoch, we numerically integrate Eq.~\eqref{eq-tot} through the matter- and dark-energy-dominated eras, neglecting the accretion of the cosmological fluid. This procedure provides a lower bound on the final mass. The corresponding solutions for each epoch are shown in Figs.~\ref{fig:Matter} and \ref{fig:DarkEnergy}. The complete mass evolution across the four cosmological epochs—namely inflation, radiation, matter, and dark energy—is displayed in Fig.~\ref{fig:History}. To conclude, a black hole with an initial mass of no more than $M(t_\mathrm i)=5.52687\times10^{-3}\,\text{g}$ at the onset of inflation would have a mass of $M=1.043\times10^{-3}\text{M}_{\odot}$ today.

\begin{table}
\centering
\begin{tabular}{ccccc}
\hline\hline\noalign{\smallskip}
Time & $t_\mathrm f$ & $t_\mathrm{rm}$ &$t_\mathrm{mde}$ & $t_0$\\
\hline
Mass $[M_\odot]$ & $5.34382\times10^{-31}$ & $2.14391\times10^{-7}$ & $7.53862\times10^{-4}$ & $1.043\times10^{-3}$
\label{table:Table2}
\end{tabular}
\caption{Mass of the black hole at the end of every epoch and today, assuming $\gamma = \gamma_3$.}
\end{table}







\section{A lower mass limit at formation time}

Now that an upper limit has been obtained for the mass of the black hole at the end of inflation, we can now proceed to determine if there is a lower limit for the mass of the black hole at the time of its formation. During inflation, if the mass is smaller than this critical value, Hawking radiation becomes so significant that the black hole evaporates before the end of inflation.

To this end, we will consider the  equation
\begin{equation}
\begin{split}
    \frac{dM}{dt}&=\gamma w\dot{a}(t)-\frac{A(M)}{M(t)^2},\\
    M(t_{\mathrm{i}})&=M_{\mathrm{i}},\\
    \gamma&=\gamma_3.
    \label{3.4}
    \end{split}
\end{equation}
where $M_{\mathrm{i}}$ denotes the initial mass of the black hole at the onset of inflation. We are solving \eqref{3.4} to determine the value of $M_{\mathrm{i}}$ such that and it is such that it does not evaporate due to Hawking that at $t = t_{\mathrm{f}}$, the mass completely vanishes (the growth of the black hole horizon due to the accelerated expansion cannot counterbalance the loss of mass due to Hawking radiation).  

\begin{figure}[t]
    \centering
    \includegraphics[width=0.7\textwidth] {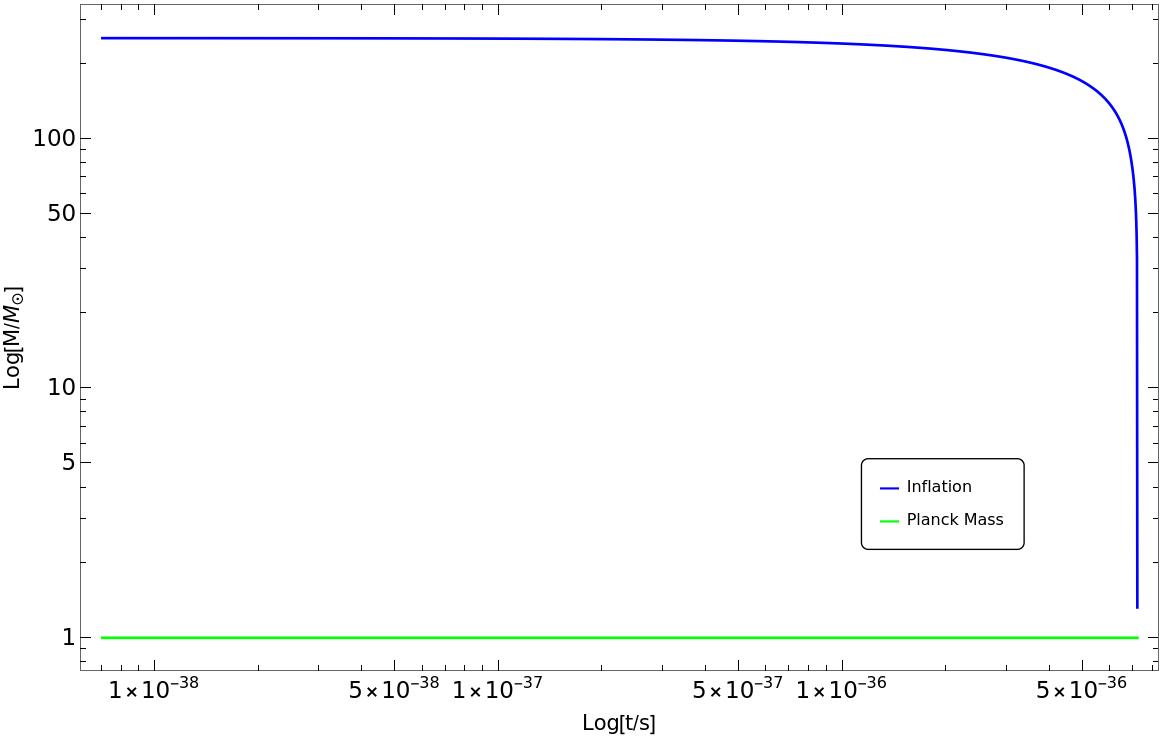}
    \caption{The mass evolution of a black hole with an initial mass of $M_\mathrm i=200 m_{\mathrm{Planck}}$ during the inflation era is shown on a logarithmic scale. As can be seen, the black hole evaporates before inflation ends at time $t=t_f$.}
  \label{fig:Planck}
\end{figure}

As shown in Figure  \ref{fig:Planck}, we obtain that for black holes such that $M_{\mathrm{i}} \le M_{\mathrm{i,min}}=253.718\,m_{\mathrm {Planck}}\simeq5.52204\times10^{-3}\, \mathrm g$, they will evaporate before inflation ends.


\section{Discussion and conclusions}

If black holes were present during the inflationary era and have survived to the present day, their masses must satisfy constraints arising from both astrophysical and cosmological considerations. In this work, we assume that these black holes are coupled to the background cosmological dynamics, so that their properties evolve in response to the changing cosmological environment.

Here we briefly summarize the constraints obtained so far:
\begin{itemize}
\item \textbf{First upper limit.} In order for the black hole event horizon to remain smaller than the particle horizon throughout all cosmological eras, the black hole mass parameter must satisfy $\gamma\leq\gamma_2=1.33324\times10^{45}$.

\item \textbf{Second upper limit.} During the radiation-dominated epoch, we consider that the black hole mass evolves due to three distinct processes: coupling to the background cosmological dynamics, mass loss via Hawking radiation, and mass gain due to radiation accretion. The latter process imposes a stringent upper bound on the black hole mass at the end of inflation, ensuring that the mass evolution remains well defined at all epochs. We find that $\gamma\leq\gamma_3=2.07348\times10^{30}$, and $M(t_{\mathrm f})\leq1062.35$ g.

\item \textbf{First lower limit.} We also derive a lower bound on the black hole mass at the onset of inflation such that the black hole does not completely evaporate due to Hawking radiation before reaching the radiation-dominated era. Accordingly, $M(t_\mathrm i)\geq M_\mathrm i=253.718\,m_{\mathrm{Planck}}$.
\end{itemize}

\textbf{Summing up:} If the gravitational mass of a black hole $M(t)$, for
$t \in [t_{\mathrm{i}}, t_{\mathrm{f}}]$, satisfies
\begin{equation}\label{ine_mass}
M(t_{\mathrm{i}})=253.718\,m_\mathrm{Planck} < M(t) < M(t_{\mathrm{f}})=1062.35\,\mathrm{g},
\end{equation}
then such black holes could still be present today. Taking into account that their masses evolve due to the coupling of their horizons to the cosmic expansion, mass loss via Hawking radiation, and mass gain during the radiation-dominated epoch, this black hole population is expected to have present-day masses within 
\begin{equation}\label{max_value}
M(t_0)\simeq 1.043\times10^{-3}\,M_\odot,
\end{equation}
which lies within the subsolar regime. This value was derived assuming $\gamma = \gamma_3 = 2.07348\times10^{30}$. In other words, black holes present during the inflationary period whose initial masses satisfy inequality \eqref{ine_mass} could have a present-day mass of at most $M(t_0)\simeq 1.043\times10^{-3}\,M_\odot$.

This mass range is subject to constraints from microlensing surveys and dynamical bounds \cite{car+21,car+26}. The results presented here should therefore be interpreted as establishing the survival window of such objects, independently of their cosmological abundance. A full assessment of their viability as dark matter candidates would require combining the mass evolution derived here with observational bounds on their present-day density fraction. Furthermore, this result depends on the specific form of the cosmological coupling and on the effective description of radiation accretion employed here, and should therefore be regarded as model-dependent rather than a generic prediction.

The result \eqref{max_value} provides an upper bound on the present-day mass of black holes existing in the inflationary period. However, for $\gamma < \gamma_3$, the present-day mass may be significantly smaller; in particular, values as low as $M(t_0)\simeq 10^{-11}\,M_\odot$ are possible. This corresponds to the asteroid-mass window, which has been discussed as a viable mass range in which primordial black holes could account for all dark matter (see, e.g., Ref.~\cite{car+21}). Such black holes would have a mass of $M(t_{\mathrm{i}}) = 598$ g at the onset of inflation.


In the standard scenario of isolated primordial black holes, objects with masses $M \lesssim 5 \times 10^{14}$ g
 are expected to have completely evaporated by the present epoch due to Hawking radiation (see e.g. Refs. \cite{car+21,car+26}). In contrast, within the cosmologically coupled framework considered here, black holes with initial masses well below this conventional evaporation threshold may survive, as the growth induced by the coupling to the background expansion counterbalances the Hawking mass loss during inflation and subsequent epochs. The survival window derived in this work therefore differs qualitatively from the standard PBH scenario.

We emphasize that the mass divergence encountered in the accretion term is not a novel feature of our analysis. The pathological behavior of the mass evolution of a Schwarzschild black hole embedded in a thermal bath was first identified by Zel'dovich and Novikov \cite{zel+67} and has been revisited and studied in depth by Barrau \emph{et al.}~\cite{bau+22}. As shown in that work, when accretion is modeled using a Schwarzschild background and a Bondi--Hoyle--type prescription, the black hole mass generically either diverges or vanishes in a finite time, even when Hawking evaporation is consistently included. Haque and collaborators \cite{haq+26} (see also \cite{kal+26}) recently analyzed the evolution of primordial black holes formed in the early universe surrounded by a thermal bath; they also found that for some critical value of the parameters of their model, the PBHs undergo runaway mass growth.


In the present work, we adopt Zel'dovich and Novikov \cite{zel+67}  effective accretion prescription as a first exploratory step to estimate the impact of radiation accretion in a cosmological setting. Our goal is not to provide a fully self-consistent description of black hole growth, but rather to identify the mass ranges for which black holes may survive the early universe under simplified assumptions. A complete and physically robust treatment would require modeling accretion onto a genuinely dynamical, cosmological black hole, consistently accounting for the relativistic nature of the horizon and the coupling to the expanding background spacetime. A possible approach to derive the mass accretion rate onto the cosmological black hole, is following the formalism introduced by Babichev et al. \cite{bab+04} (see also \cite{ka+25}); it consists of the integration of the time component of the energy-momentum conservation law associated to the cosmological black holes solution. A more complete treatment will be presented elsewhere.


Our analysis determines the allowed mass range for black holes present during inflation to survive until today under the dynamical evolution considered here. We do not attempt to model their formation process or compute their cosmological abundance. The results therefore identify the mass interval compatible with cosmological expansion, Hawking evaporation, and radiation accretion, rather than establishing any specific contribution to the dark matter density.

The mechanism explored in this work shows that coupling to the background expansion can substantially modify the standard picture of primordial black hole evolution. A fully self-consistent treatment, in particular, one incorporating a relativistic description of accretion onto dynamical cosmological black holes, will be required to assess the quantitative implications for present-day black hole populations and their possible cosmological role. These issues will be addressed in future work.







\section*{Declarations}


\begin{itemize}

\item Acknowledgments
D.P. and M.E.U. are grateful to Gustavo E. Romero and Santiago E. Perez Bergliaffa for their valuable and insightful comments on this manuscript.

\item Funding
D. P. acknowledges the support from CONICET under Grant No. PIP 0554 and AGENCIA I+D+i under Grant PICT-2021-I-INVI- 00387.

\item Competing interests 

\item Ethics approval 

\item Consent to participate

\item Consent for publication

\item Availability of data and materials

\item Code availability

\item Authors' contributions All authors contributed equally to the manuscript.

\end{itemize}






\input{sn-article.bbl}


\end{document}

%% file: sn-article.bbl
\providecommand{\noopsort}[1]{}\providecommand{\singleletter}[1]{#1}%